\documentclass[thmsa]{article}
\usepackage{amssymb}

\input{tcilatex}
\begin{document}

\author{Babur M. Mirza\thanks{%
Email address: bmmirza2002@yahoo.com} \\
Department of Mathematics, \\
Quaid-i-Azam University, \\
Islamabad 45320, Pakistan}
\title{GRAVITOMAGNETIC RESONANCE SHIFT DUE TO A SLOWLY ROTATING COMPACT STAR}
\date{August 4, 2003}
\maketitle

\begin{abstract}
The effect of a slowly rotating mass on a forced harmonic oscillator with
two degrees of freedom is studied in the weak field approximation. It is
found that according to the general theory of relativity there is a shift in
the resonat frequency of the oscillator which depends on the density and
rotational frequency of the gravitational source. The proposed shift is
quite small under normal physical situations however it is estimated that
for compact x-ray sources such as white dwarfs, pulsars, and neutron stars
the shift is quite appreciable.
\end{abstract}

\section{\protect\vspace{0in}Introduction}

The general theory of relativity is logically one of the most compelling
theories of modern physical science. However being a physical theory it is
imperative that it should not only describe as many different physical
phenomenon as possible but also make accurate predictions of physical
quantities relevant to observations. As regards the accuracy in prediction
the general theory of relativity has done extremely well in all its known
tests such as the perihelion shift of mercury, deflection of light, and
gravitational redshift experiments$^{1,2}$. However the range of
applicability where general theory of relativity can be compared with direct
observations has remained small, owing to the fact that under normal
physical situations on earth and inside the solar system it is very
difficult to differentiate its predictions from other rival theories such as
the Newton's theory of gravitation, even when it is considered under its
linear approximation. The discovery of compact x-ray sources in the universe
has opened up new avenues to the test general relativity in naturally
existing systems. In these astrophysical laboratories extreme physical
conditions are encountered which are impossible to be reproduced on earth.
For example$^{3}$ densities of a usual compact x-ray source is of the order
up to $10^{28}$ $kg/m^{3}$ and rotational periods $10^{10}Hz$. Under such
conditions general relativity provides not only a consistent picture of
various physical aspects of the phenomenon but also predicts, with
unprecedented accuracy, magnitudes of various physical parameters involved
such as the timing of rotational periods and their decay with time$^{4}$. On
the other hand all such systems are composed of atoms existing under a very
strong gravitational field. Within the classical theory these atoms can be
taken as forced harmonic oscillators whose physical behavior depends very
much on the background gravitational field.

In this paper we study the effects of gravitational field on the behavior of
such oscillators within the framework of the general theory of relativity..
After giving a preliminary introduction, we write the equations of motion
for a forced harmonic oscillator in section 2, making a plausible physical
assumption that the star is rotating slowly (i.e., far below the speed of
light limit). In section 3 we discuss the solution to the equations of
motion and especially their behavior near the resonant frequency. In section
4, we give the behavior of the oscillators for compact astrophysical objects
and also give a comparison. Lastly a brief summary and conclusion is given
in section 5.

\section{The Equations of Motion}

\subsection{\textit{Preliminaries}}

According to the general theory of relativity$^{5,6}$ the motion of a free
particle of gravitational mass $m_{g}$ is given by the geodesic equation

\begin{equation}
\frac{d^{2}x^{\alpha }}{ds^{2}}+\Gamma _{\beta \gamma }^{\alpha }\frac{%
dx^{\beta }}{ds}\frac{dx^{\gamma }}{ds}=0;\quad \alpha ,\beta ,\gamma
=0,1,2,3  \label{1}
\end{equation}
where $x^{\alpha }$ $\equiv (\mathbf{x}^{i},x^{0})=(\mathbf{r,}t)$ are the
position coordinates of the test particle and $i=1,2,3.$ Here $s$ is the
affine parameter and $\Gamma _{\beta \gamma }^{\alpha }=g^{\delta \alpha
}(g_{\delta \beta ,\gamma }+g_{\delta \gamma ,\beta }-g_{\beta \gamma
,\delta })/2$ is the Christoffel symbol, where $g^{\alpha \beta }$ is the
metric tensor and `$,_{\alpha }$' denotes differentiation with respect to $%
x^{\alpha }$.

Let the metric tensor be expressed as

\begin{equation}
g_{\alpha \beta }=\eta _{\alpha \beta }+h_{\alpha \beta }  \label{2}
\end{equation}
where $\eta _{\alpha \beta }$ $=diag(-1,-1,-1,1)$ is the Minkowski metric
tensor and $h_{\alpha \beta }$ is the perturbation to the metric. Then under
the linear approximation $h_{\alpha \beta }\ll 1$ we can take time $t$ as
the affine parameter. Requiring rotation to be slow (i.e., neglecting square
and higher powers of velocity vector $\mathbf{v\equiv }d\mathbf{r}/dt$) we
obtain the following expression for acceleration for a slowly rotating
sphere of homogeneous mass density

\begin{equation}
\frac{d^{2}\mathbf{r}}{dt^{2}}=\mathbf{G}+\mathbf{v}\times \mathbf{H}
\label{3}
\end{equation}
where

\begin{equation}
\mathbf{G=-\nabla }\varphi ,\quad \mathbf{H=\nabla \times }4\mathbf{a}
\label{4}
\end{equation}
and

\begin{equation}
\varphi =-\iiint \frac{\rho }{r}dV,\quad \mathbf{a=}\iiint \frac{\rho 
\mathbf{v}}{r}dV  \label{5}
\end{equation}
$\rho $ being the mass density and $V$ is the volume. Here the remarkable
formal analogy of the above results with the classical electromagnetic
theory must be noticed with the \textit{proviso }that a factor of 4 is to be
multiplied with the gravitomagnetic vector potential $\mathbf{a}$. Pursuing
this analogy, we find$^{7}$ that for a slowly rotating sphere of homogeneous
mass density first term in expression (3) is the gravitoelectric (GE)
potential given by the Newtonian gravitational potential $-M\hat{\mathbf{r}}%
/r$ (where $M$ is the mass of the gravitating source, and $G=1=c$ unless
mentioned otherwise). Moreover there is the gravitomagnetic (GM) potential $%
\mathbf{v}\times \mathbf{H}$ where

\begin{equation}
\mathbf{H}=-\frac{12}{5}MR^{2}(\mathbf{\Omega .r}\frac{\mathbf{r}}{r^{5}}-%
\frac{1}{3}\frac{\mathbf{\Omega }}{r^{3}})  \label{6}
\end{equation}
$R$ being the radius and $\mathbf{\Omega }$ is the angular velocity of the
gravitational source. The second part of equation (3) has a non-Newtonian
origin. It is regarded to exhibit typically Einsteinian effects and
therefore plays an important role in testing Einstein's theory of
gravitation in the weak field and slow rotation approximation$^{8,9}$. The
independence of the GM potential from a particular frame and from a
particular coordinate system used has been demonstrated thus making it
physically significant$^{10}$. This effect can be interpreted as
`gravitomagnetic current' induced in the vicinity of the gravitational
source due to its rotation. For its analogy with the Lorentz force law for
the electromagnetic field the force $m_{g}(\mathbf{G}+\mathbf{v}\times 
\mathbf{H)}$ is often referred to as gravitoelectromagnetic (GEM) force.
However to measure the effects of this force on a given physical system high
accuracy in experiments is required where it is necessary to isolate the
experimental setup from contingencies$^{11,12}$.

\subsection{\textit{Forced Harmonic Oscillations in Weak Gravitational Field
and Slow Rotation Approximation}}

Let us consider a forced damped harmonic oscillator having two degrees of
freedom placed in a GEM field. If $m_{I}$ is the inertial mass of the test
particle then in addition to the binding force $-m_{I}\omega _{0}^{2}\mathbf{%
r}$ , the damping force $-m_{I}\gamma \mathbf{v}$, and an external force $%
F_{0}\exp i\omega t$ acting along, say, the x direction, there also acts on
the particle a GEM force $m_{g}(\mathbf{G}+\mathbf{v}\times \mathbf{H})$.
Considering the GE field of magnitude $-g$ to be along y-axis and the GM
field of magnitude $H$ in the direction normal to the xy-plane$^{13}$, we
find that the GEM force has a component $m_{g}Hv_{y}$ in the x direction and
a component $-m_{g}Hv_{x}$ in the y direction; where $v_{x}$ and $v_{y}$ are
the components of velocity in the x and y directions respectively. With
these additional terms and the fact that $m_{g}=m_{I}(=m)$, in accordance
with the general theory of relativity, we get the equations of motion as a
coupled system of ordinary differential equations

\begin{equation}
\frac{d^{2}x}{dt^{2}}=-\omega _{0}^{2}x-\gamma \frac{dx}{dt}+\frac{F_{0}}{m}%
\exp i\omega t+H\frac{dy}{dt},  \label{7}
\end{equation}

\begin{equation}
\frac{d^{2}y}{dt^{2}}=-\omega _{0}^{2}y-\gamma \frac{dy}{dt}-g-H\frac{dx}{dt}
\label{8}
\end{equation}

Now let us introduce the auxiliary variable $\Re =x+iy$, then the equations
of motion give a single equation

\begin{equation}
\frac{d^{2}\Re }{dt^{2}}+(\gamma +iH)\frac{d\Re }{dt}+\omega _{0}^{2}\Re =%
\frac{F_{0}}{m}\exp i\omega t-ig  \label{9}
\end{equation}

Notice that the GM parameter $H$ has a coupling to the damping parameter $%
\gamma $ in the complex plane ($x,y$). Dimensional analysis of equations (7)
and (8) shows that $H$ must have dimensions of cycles per second i.e., of
frequency. These considerations give rise to the question whether these
`gravitomagnetic oscillations' produce significant effects on the physical
behavior of an oscillatory system. As we shall see in the following that the
answer to this question must be given in affirmative. However these effects
are extremely small except in those cases where density and rotational
frequency of the gravitational source are extremely large, such as those
found in compact astrophysical x-ray sources.

\section{Solution to the Equations of Motion and Interpretation}

The exact solution to equation (9) can be obtained, however we are
interested in studying the behavior of amplitude especially near the
resonant frequency. To do so we express $\Re $ as $A\exp i\omega t$ where $A$
is the complex amplitude$.$ Substitution in equation (9) gives

\begin{equation}
A(\omega )=\frac{\frac{F_{0}}{m}-ig\exp (-i\omega t)}{\omega _{0}^{2}-\omega
^{2}-\omega H+i\omega \gamma }  \label{10}
\end{equation}

Multiplying the last expression with its complex conjugate and taking square
root we have the following real expression for the amplitude

\begin{equation}
\left| A(\omega )\right| =\sqrt{\frac{(F_{0}/m)^{2}+g^{2}-2(F_{0}/m)g\sin
\omega t}{(\omega _{0}^{2}-\omega ^{2}-\omega H)^{2}+\omega ^{2}\gamma ^{2}}}
\label{11}
\end{equation}

Comparing this expression for the amplitude with the expression for the
amplitude of a forced damped harmonic oscillator having two degrees of
freedom we notice that there are two main differences, one the addition of $%
g $ dependent terms in the numerator, two an addition of the term $-\omega H$%
, in the denominator, to the square of eigenfrequency $\omega _{0}$ of the
system. The second difference is of particular importance since it involves
the effect of GM field on the resonant frequency of the system. To estimate
this effect we notice that in equation (11) for small damping i.e., $\gamma
\ll \omega $ and given values of the parameters $F_{0}/m$ $,g,$ and $\omega
_{0}$, the amplitude $A$ is maximum when $\omega _{0}^{2}-\omega ^{2}-\omega
H\simeq 0$ i.e., when the applied frequency is approximately $(-H\pm \sqrt{%
H^{2}+4\omega _{0}^{2}})/2.$ Expanding in powers of $(H/2\omega _{0})^{2}$
and requiring that $H\ll \omega _{0}$ we obtain the following approximate
result for resonant frequency

\begin{equation}
\omega \simeq \omega _{0}-\frac{H}{2},\qquad \gamma \ll 1,\quad H\ll \omega
_{0}  \label{12}
\end{equation}
This result shows that there is a shift in the resonant frequency by an
amount $H/2$.

\section{Application to Compact Stars}

The outer atmosphere of a white dwarf consists mainly of fully ionized $%
^{4}He$, $^{12}C$, and $^{16}O$ atoms forming a crystalline structure in
which electrons can be taken as forced harmonic oscillators, moving under
the influence of lattice vibrations$^{14}$. Along the radial direction the
motion of these electrons is very much limited by the GE attraction and the
Fermi pressure, therefore in any thin layer enveloping the star the
electrons can be taken as oscillating with two degrees of freedom. As the
star of mass greater than the Chandrasekhar limit exhausts its thermonuclear
fuel the GE force not only contracts the core (increasing the mass density
per unit area), its angular frequency also is increased from $1Hz$ $(=1$
cycle per second$)$ to $1MHz$ or above. To study the resonant behavior of
these electrons we begin with relatively less severe case of white drawf of
mass $1M_{\circledast }=1.989\times 10^{30}kg$ and radius $7\times 10^{6}m$
rotating at frequency $1Hz$ . Then the case of the pulsar of mass $%
1.4M_{\circledast }$ and radius $3\times 10^{4}m$ rotating at frequency $%
30Hz $ is studied. Finally we consider a neutron star of mass $%
2M_{\circledast }$and radius $1\times 10^{4}m$ rotating at frequency $1kHz$.
In these cases $g$ $(=GM/R^{2}$ , where $G=6.67\times 10^{-11}Nm^{2}/kg^{2})$
ranges from $2.7075\times 10^{6}m/\sec ^{2}$ (for the white dwarf) to $%
2.0636\times 10^{11}m/\sec ^{2}$ (for the pulsar) to $2.6533\times
10^{12}m/\sec ^{2}$ (for the neutron star); consequently $%
(F_{0}/m)^{2}+g^{2}+2(F_{0}/m)g\sin \omega t$ varies from order $%
10^{12}m/\sec ^{2}$ to $10^{24}m/\sec ^{2}$ where $\theta =\omega t$ is the
phase. Now from equation (6) we have in the equatorial plane at the surface
of the star i. e., $r=R$ :

\begin{equation}
H\equiv \mid \mathbf{H\mid =}\mu \frac{M}{R}\mid \mathbf{\Omega }\mid 
\label{13}
\end{equation}
where $\mu =4G/5c^{2}\cong 0.5940\times 10^{-27}m/kg$ . For the above given
values of the relevant physical parameters we have from expression (13), $H$
is $0.1687\times 10^{-3}Hz$ , $1.6540Hz$, and $236.2932Hz$ for the
whitedwarf, the pulsar and the neutron star respectively. With these values
of parameters we plot the resonance curves, using expression (11), for the
three cases in figure (1) and (2) for electron with eigenfrequency $1kHz$ , $%
\theta =\theta _{1}=0,$ and $\gamma \approx 0.1Hz$.

We notice that for a slow rotating pulsar the resonance shift $\Delta \omega
\equiv \mid \omega -\omega _{0}\mid =H/2$ is not very large (about $0.8269Hz$%
) as compared with the shift for the case of the neutron star (about $%
117.3196Hz$).

\section{Summary and Conclusion}

The above considerations can be summed up as follows: \textit{under the
influence of gravitomagnetic field of a rotating compact star there is a
shift in the resonant frequency of an oscillator lying in its vicinity. The
shift depends on the density and on the rotational frequency of the star. }

Inside the solar system and especially in terresrial laboratories the
detection of gravitomagnetic resonance shift is very unlikely (for Earth $H$
is approximately $0.6440\times 10^{-14}Hz$ and for the sun it is $%
0.6597\times 10^{-13}Hz$). However for compact gravitational objects the
shift lies in the domain of observations. These observations may not be
directly possible at present, but with further theoretical as well as
observational developments, it may well become possible to provide evidence
for such an effect.

\textbf{Acknowledgments} I am grateful to to Dr. B. J. Ahmedov for bringing
to my attention reference (13), and for various comments. I also thank Dr.
A. Qadir for useful comments.

\subsection{Figure Captions}

Figure 1: Amplitude plots for an oscillating test particle exhibiting the
resonance shift for a white dwarf (mass $1M_{\circledast }$ , radius $%
7\times 10^{6}m$ , rotational frequency $1Hz$)and a pulsar (mass $%
1.4M_{\circledast }$, radius $3\times 10^{4}m$ , rotational frequency $30Hz$%
) as the gravitational source. The frequency $\omega $ is measured in $Hz$
and the amplitude is in units $m/Hz.\sec ^{2}$.

Figure 2: Amplitude plots for an oscillating test particle exhibiting the
resonance shift for a pulsar (mass $1.4M_{\circledast }$, radius $3\times
10^{4}m$ , rotational frequency $30Hz$) and a neutron star (mass $%
2M_{\circledast }$ , radius $10^{4}m$ , rotational frequency $1kHz$) as the
gravitational source. The frequency $\omega $ is measured in $Hz$ and the
amplitude is in units $m/Hz.\sec ^{2}$.

\end{document}